# Three-Dimensional Magnetic Page Memory

O. Ozatay[1*], A. Gokce[1], T. Hauet[2], L. Folks[3], A. Giordano[4], G. Finocchio[4]

[1] *Physics Department, Bogazici University, Bebek 34342, Istanbul, TURKEY*
[2] *Institut Jean Lamour, UMR CNRS-Université de Lorraine, 54506 Vandoeuvrelès Nancy, FRANCE*
[3] *Department of Electrical Engineering, School of Engineering and Applied Sciences, University at Buffalo, NY 14260, USA*
[4] *Department of Mathematical and Computer Sciences, Physical Sciences and Earth Sciences, University of Messina, Messina, ITALY*

## *ABSTRACT*

The increasing need to store large amounts of information with an ultra-dense, reliable, low power and low cost memory device is driving aggressive efforts to improve upon current perpendicular magnetic recording technology. However, the difficulties in fabricating small grain recording media while maintaining thermal stability and a high signal-to-noise ratio motivate development of alternative methods, such as the patterning of magnetic nano-islands and utilizing energy-assist for future applications. In addition, both from sensor and memory perspective three-dimensional spintronic devices are highly desirable to overcome the restrictions on the functionality in the planar structures. Here we demonstrate a three-dimensional magnetic-memory (magnetic page memory) based on thermally assisted and stray-field induced transfer of domains in a vertical stack of magnetic nanowires with perpendicular anisotropy. Using spin-torque induced domain shifting in such a device with periodic pinning sites provides additional degrees of freedom by allowing lateral information flow to realize truly three-dimensional integration.

*\* Correspondence should be addressed to: ozhan.ozatay@boun.edu.tr*

The application of instantaneous heat pulses to achieve brief periods of temperature increase in a recording medium results in sudden changes in a number of physical properties that can be exploited in a wide spectrum of data storage applications. This approach is already used in optical recording such as CDs (compact discs) and DVDs (digital versatile discs) where the heat source is a laser pulse that causes a change in the crystallinity of the chalcogenide recording medium leading to marked modifications of the optical reflectivity[1]. If the laser heat source is replaced by resistive heating from electrical pulses and the physical property being monitored is the electrical resistivity of the chalcogenide material then the resulting device (phase-change memory) is now being utilized in storage-class memory applications[2,3]. In magnetic recording, in order to overcome the limitations in scaling imposed by the thermal stability of small grains[4], thermally-assisted recording with a laser beam coupled to a near field transducer as the heat source instantaneously reduces the reversal fields required to write high-anisotropy media, allowing thermally stable grain sizes down to a few nm[5,6]. Even though the local injection of electrical current pulses using a scanning tunneling microscope tip has been shown as an effective way to induce local heating to allow the manipulation of magnetization at the nanoscale[7], a magnetic memory device concept taking advantage of such local heating due to electrical pulses has not yet been realized.

Here we utilize the local heating from electrical pulses in a three dimensional magnetic memory device[8] (magnetic page memory) that comprises an array of magnetic nanowires with perpendicular magnetic anisotropy, stacked in a cross-wire configuration where individual layers of nanowires are electrically isolated with a low-thermal conductivity silicon-nitride spacer[9] as shown in Fig. 1a-b. Such a memory design with three dimensional storage capability is very desirable from a technology point of view[10,11,12] since it is expected to provide a low-cost per bit, high-speed, and ultra-dense memory solution. In addition, such a three dimensional magnetic memory does not suffer from loss of information in the long run due to resistance drift in PCM[3] and charge leakage in flash memories. This high potential inspired the industry to investigate alternative approaches such as three dimensional magnetic domain-wall racetrack memory[13,14,15] based on the interaction of a spin-polarized current pulse with a magnetic domain-wall[16]. However the nano-fabrication challenges involved in the actual realization of an array of vertical race-tracks with periodic pinning sites required for the three-dimensional geometry present significant obstacles[14].

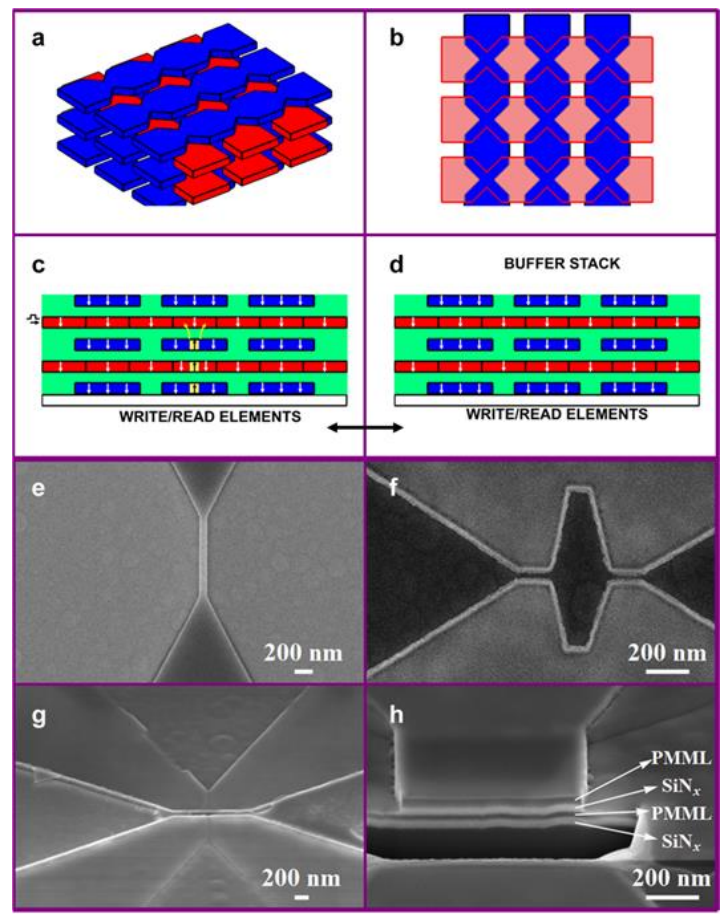

**Figure 1:** **Magnetic Page Memory - Principle of Operation.**

**a,** Three-dimensional schematics of the perspective view of an array of nominally identical, perpendicular anisotropy magnetic nanowires with periodic nucleation/pinning sites stacked in a cross-wire configuration. Alternating layer of nanowires are colored blue and red.

**b,** The top view schematics of the nanowire stack displaying the vertical alignment of pinning sites acting as storage bits in the cross-wire architecture. Alternating layer of nanowires are colored blue and red.
**c, d** The side view schematic drawing of the page memory device with a single layer of read/write elements, a stack of nanowire arrays with perpendicular magnetization separated by a silicon-nitride low thermal conductivity spacer (green region) where the straight arrows represent the local magnetization direction and the curved arrows represent the local stray field. A square electrical pulse is applied to the layer to be written to instantaneously lower its coercivity. A buffer stack in d is a replica of the memory array that is used for temporary storage in the read process.
Scanning electron microscopy images of CoNi-Pd multilayer nanowires:
**e,** Single nanowire with 100 nm width and 1 μm length, designed for writing experiments
**f,** A nanowire with a double-constriction designed for spin-torque induced lateral domain transfer studies.
**g,** Two nanowires with a 20 nm thick silicon-nitride spacer in the cross-wire configuration designed to study the heat-assisted vertical domain transfer in page memory.
**h,** A focused-ion-beam cut of the sample shown in (g) showing the film structure consisting of perpendicular anisotropy magnetic multilayers (PMML) separated by silicon-nitride spacer.

## **Operation Principle**

The basic operation principle of a magnetic page memory device is shown in Fig 1a-d. A perpendicular anisotropy magnetic nanowire array with periodic domain nucleation sites (constrictions) in a cross-wire architecture is built on a single layer of read/write elements. After all the data have been recorded on the first magnetic layer in the form of the presence or absence of domains in the nanowire constrictions, the domain configuration is replicated to the next layer by instantaneously reducing the nucleation fields through local Joule heating from electrical pulses. The increasing temperature in the nucleation sites of the second layer makes its local magnetization susceptible to the stray fields coming from the recorded domains in the first magnetic layer. The third magnetic layer constrictions are also heated synchronously with the second magnetic layer so that the stray field coming from the top layer is suppressed. After the electrical pulses are turned off, the second layer domain configuration ends up being a replica of the first magnetic layer. The shift of the information recorded on the second magnetic layer to the layer above it proceeds in a similar manner through sequential heating of the third and fourth magnetic layers and cooling in the presence of the stray fields available from the perpendicularly oriented domains as shown in Fig. 1c.

To reach a data point recorded on a given layer, a buffer stack (Fig. 1d) is used to maintain the information stored beneath the layer of interest. For instance, to read data from the second layer, the information on the first layer is read and transferred by the read/write elements to the first layer of a buffer stack. The read element can for instance be a conventional tunneling-magnetoresistance (TMR) head and the write element could be a conventional perpendicular write head already in use in perpendicular magnetic recording heads[8]. Then by applying electrical pulses in the same manner as above the first layer replicates the domain structure of the second layer making the data accessible to the read/write elements. Here the insulating spacer thickness is chosen to ensure both magnetic transparency and thermal isolation between layers to avoid disturbing the magnetization configuration of neighboring layers (see Supplementary Note1). The periodic constrictions where the domains nucleate also act as pinning sites both to fix the position of the information (domains) and to enable highly concentrated current densities that give rise to the localized heating process for the vertical domain transfer process.

**Experimental Setup**

The different device types we have studied to demonstrate the page memory concept, are summarized in Fig. 1e-h. We have used magnetic nanowires with a composition of Ta (1.5)/Pd (3)/ [$Co_{55}Ni_{45}$ (0.22) / Pd (1.2)] x 22 repeats /Pd (2) where all the thicknesses are in nanometers. The single layer-single nanowire structures (Fig. 1e) served to characterize the writing conditions (see Supplementary Method1). The double constriction samples (Fig. 1f) enabled the study of lateral domain transfer by using spin transfer torque[17]. In a double layer version (Fig. 1g) the individual layers of nanowires in the cross-wire architecture were isolated with a 20,40 or 60 nm thick silicon-nitride spacer as the low thermal conductivity spacer. Fig. 1h shows a scanning electron microscope (SEM) image of a focused ion beam (FIB) cut of the double-layer structure in Fig. 1g. The spacer layer sandwiched in between perpendicular anisotropy magnetic multilayer (PMML) nanowires plays several crucial roles in the operation of such a device. It not only isolates the individual layers electrically and thermally, but also ensures magnetic transparency which can be controlled with a proper choice of the film thickness.

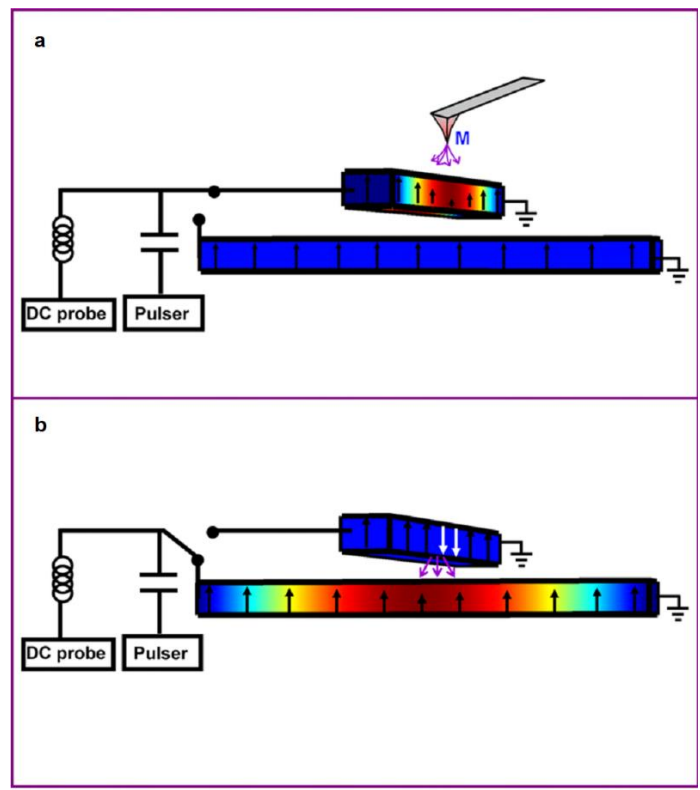

**Figure 2: Schematics of the experimental setup.**

**a,** Heat-assisted writing using a magnetic force microscopy tip placed a few nm above the surface of the top nanowire as the source of stray fields. The bottom nanowire is grounded whereas the top nanowire resistance is monitored by a dc probe before and after the application of electrical pulses.
**b,** The top nanowire reversal domain acts as the source of stray fields to write onto the bottom nanowire which is made susceptible to such fields through resistive heating from electrical pulses. The dc probe monitors the resistance of the bottom nanowire before and after the application of pulses whereas the top nanowire is grounded.

A simple schematic drawing of the experimental setup is shown in Fig. 2. For the purpose of this experiment a commercial magnetic force microscopy tip (Bruker-MESP-HM) placed a few nanometers above the surface of the top nanowire is used as the source of stray fields to induce magnetization reversal locally in the nanowire underneath. At the tip of the cantilever the magnetization is forced to point out of plane due to the shape anisotropy and stray fields as high as 500-700 Oe are available from such tips[18]. The measured bulk nucleation field of our film structure is ~3 kOe much higher than the stray fields available from MESP-HM magnetic tips. Therefore, at room temperature the magnetic tip is not able to trigger magnetization reversal. However, if an electrical pulse sufficiently high in amplitude (~3x10$^7$ A/cm$^2$) and width (~8ns) is applied to the nanowire, the local nucleation field is reduced due to increasing temperature from current-induced resistive heating (see Supplementary Note2). By grounding the bottom nanowire during the pulse application, only the top nanowire is made to be susceptible to the stray field-induced reversal as shown in Fig. 2a. A dc probe independently (through a bias tee) monitors the dc resistance of the top nanowire to detect magnetoresistance effects due to domain nucleation. To transfer the top domain to the bottom nanowire the measurement setup is reconnected to the bottom nanowire and the perpendicularly oriented domain at the top nanowire is used as a source of the stray field (the magnetic tip is removed). Analogous to the former case the bottom nanowire is heated with an electrical pulse and its magnetization is made susceptible to reversal due to the stray field (Fig. 2 b).

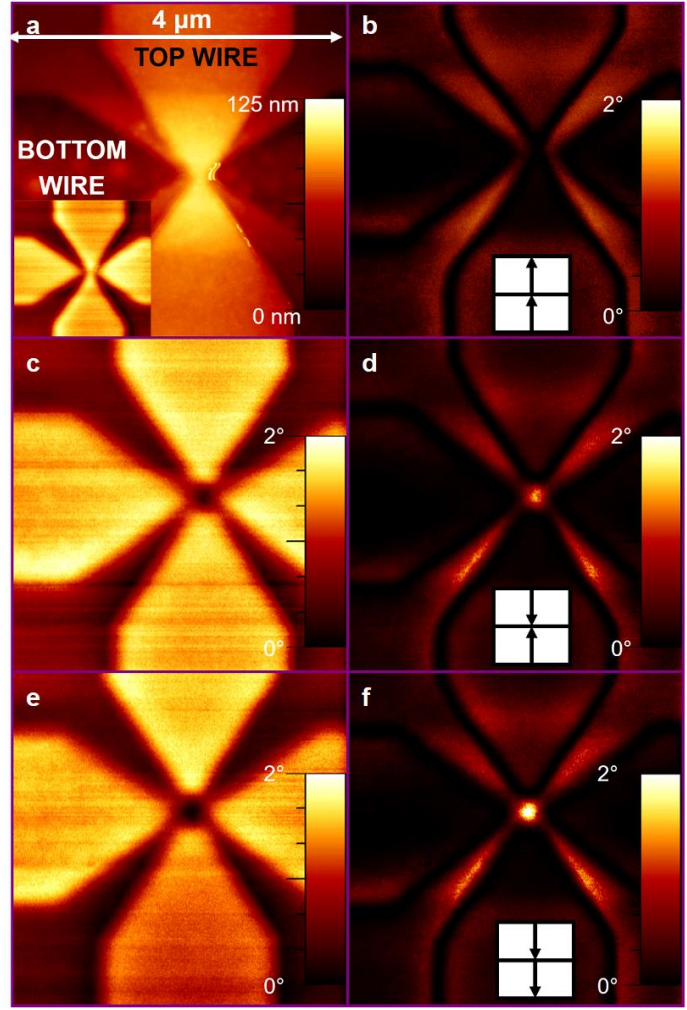

**Figure 3: Experimental detection of the vertical heat-assisted domain transfer.**

**a,** The atomic microscopy image of the topography. Inset: Magnetic force microscopy (MFM) image of the out-of-plane remanent state after saturation in a 2T out-of-plane magnetic field in a single 100nm wide constriction sample.
**b,** The MFM image in the inset of a multiplied by itself (all the pixel intensities are squared) and replotted in the same color scale as d and f. Top and bottom constriction magnetizations pointing out of plane.
**c,** MFM image after the application of a 500mV 10ns pulse to the top constriction in the presence of a magnetic tip parked 1nm above the surface.
**d,** The MFM image in c multiplied by itself (all the pixel intensities are squared) and replotted in the same color scale as b and f. Top constriction magnetization reversed.
**e,** MFM image after the application of a 500mV 10ns pulse to the bottom constriction in the presence of the stray field from the top nanoconstriction.
**f,** The MFM image in e multiplied by itself (all the pixel intensities are squared) and replotted in the same color scale as b and d. Bottom constriction magnetization reversed.

**Results**

The domain configurations for the remnant state (after the application of a 2T out-of-plane field) (Fig. 3a-inset), the state after magnetic writing (with a 500mV -10 ns wide pulse) to the top nanowire with a magnetic force microscopy (MFM) tip (Fig. 3c) and the state after domain replication process from the top to the bottom nanowire (Fig. 3e) can be directly imaged with magnetic force microscopy, provided that the magnetic signal from the bottom nanowire has a significant contribution to the image contrast. Fig 3a shows the corresponding topography image. The bottom wire lies horizontally whereas the top wire lies vertically. The 100 nm wide constrictions acting as pinning sites are aligned on top of each other. To distinguish the two cases when there is a reversal domain at the top nanowire and in both nanowires, the images shown in Fig. 3a-inset, c and e are multiplied by themselves and replotted with inverted colors as shown in Figs. 3b, d and f respectively (i.e. the MFM signal intensity in each pixel is squared). When the resulting images are plotted on the same color scale with inverted contrast, the relative contribution to the MFM signal from the bottom nanowire is visually amplified (appears brighter as in Fig. 3f). We find that this enhancement of the MFM signal (~30%) is also accompanied by a reversible change in the resistance (~90 Ω) of the bottom nanowire (~90 mΩ ≲ 0.1%) as detected by the dc probe. Such small magnetoresistance effects due to constricted domain walls have indeed been observed previously[19] and have been ascribed to domain-wall anisotropic magnetoresistance[20]. This writing process can be clearly understood from macrospin simulations based on Landau-Lifshitz-Bloch equation[21],[22]. Fig. 4a shows an example of time domain evolution of the three components of the magnetization ($m_X$, $m_Y$ and $m_Z$) for a thermally assisted stray field induced writing as a response to a thermal pulse (420 K) and for a bias field of $H_{appl} = -30$mT. This pulse is due to the Joule heating induced by the current flowing shown in Fig. 4b while the dipolar interactions originate from the bottom ferromagnetic layer, considered with the magnetization vector pointing along the - z direction. (see Supplementary Method2).

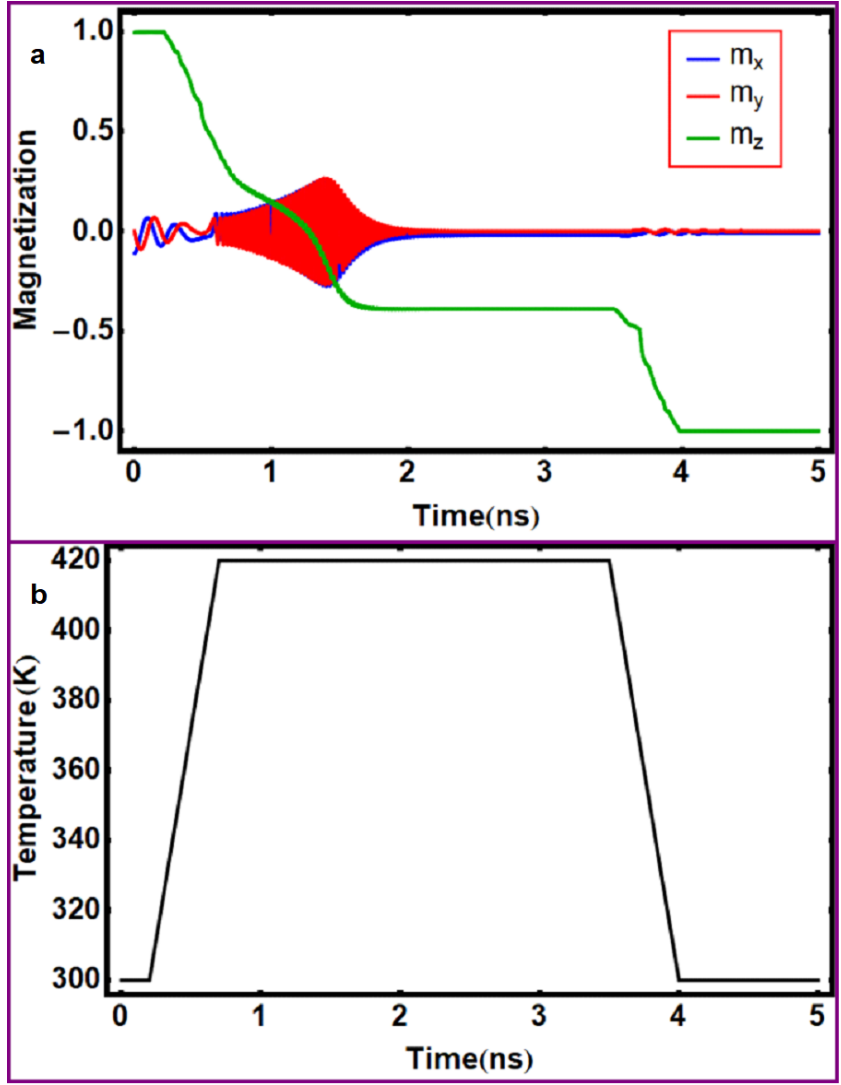

**Figure 4: Macrospin Landau-Lifshitz Bloch Simulation Results**

**a,** Time domain traces of the magnetization switching ($m_z$ switches from +1 to -1 - black line) computed with the LLB equation within the macrospin approximation as a response to the thermal pulse.
**b,** Temperature change with time. The ferromagnet is biased with an external field that is of the same order of the dipolar field that can be originated from the bottom layer (-30mT).

The ability to perform spin polarized current induced, intra-layer horizontal translation of domains, in addition to the vertical inter-layer replication, in a page memory device would enable truly three dimensional dynamic information flow. Fig. 5 shows the possible outcomes of the interaction between a spin-polarized current pulse and a written domain in a double constriction sample (with 200 nm and 100 nm wide constrictions). After saturating the magnetization out-of-plane with a 2T magnetic field Fig. 5b, at remanence we apply an 800mV amplitude, 10 ns width pulse when a magnetic tip is parked 1nm above the surface of the top constriction. Fig. 5c shows that the top constriction magnetization is reversed as a result of the local heating and the stray field from the tip. When the tip is removed and an additional electrical pulse (with electron flow from top to bottom) of 2.2V amplitude and 1ns width is applied (corresponding to $1\times10^8$ A/cm$^2$) we find that the top domain is replicated at the bottom constriction without the reversal of the intermediate region between the two constrictions (Fig. 5d). However, if the same experiment is performed with 1V-1ns ($4.5\times10^7$ A/cm$^2$) pulses, then the written domain grows into the intermediate region between the constrictions until it is finally captured at the bottom constriction after the application of 12 pulses (Fig. 5e-h). The gradual propagation of the domain wall into the spacer does not appear to be isotropic (Fig. 5g) and finds stable pinning points as the domain wall hits the hexagonal spacer region corners.

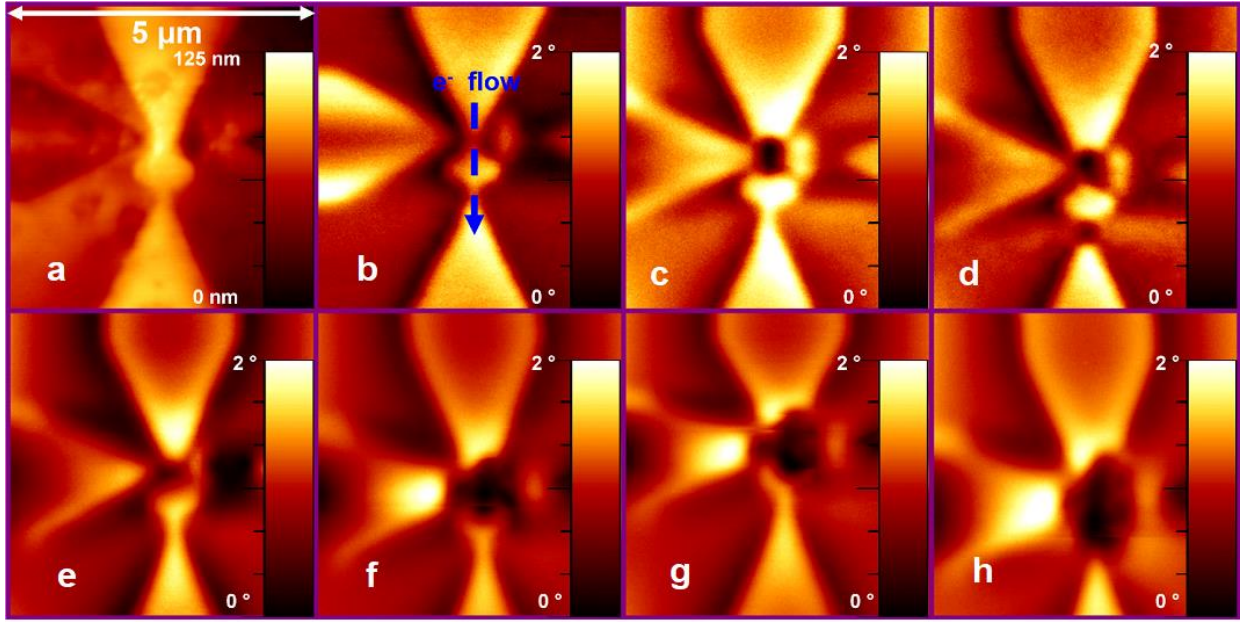

**Figure 5: Experimental detection of the spin-torque induced lateral domain transfer.**

**a,** Atomic force microscopy image of a double-constriction (200nm top-100nm bottom) sample with 500nm lateral spacing between the constrictions.
**b,** MFM image of the remanent state after saturation with a 2T out-of-plane field. Blue arrow represents the electron flow direction for pulsing experiments.
**c,** MFM image after the application of a 800 mV, 10ns pulse with a magnetic tip placed 1nm above the surface of the top constriction.
**d,** MFM image after the application of a 2.2V 1ns pulse in the absence of the magnetic tip
**e,** MFM image after resaturation and rewriting of the reversal domain in the top constriction
**f,** MFM image after the application of 1V, 1ns pulses five times in the absence of the magnetic tip.
**g,** MFM image after the application of 2 additional 1V, 1ns pulses in the absence of the magnetic tip.
**h,** MFM image after the application of 5 additional 1V, 1ns pulses in the absence of the magnetic tip.

These two distinct regimes of the domain wall - spin polarized current interaction in this system can be understood with micro-magnetic modeling (see Supplementary Method3) taking into account the spatial variations of magnetic properties due to current-induced spatially inhomogeneous heating (for details of the simulation parameters see Supplementary Note3). In these simulations we assume that the primary effect of the different temperatures (in addition to change of anisotropy and magnetization) in the two cases shown in Fig. 5 is to change the effective anisotropy direction so that lower current densities result in more out-of-plane magnetization whereas high-current densities cause enough heating to have shape anisotropy pull the effective magnetization direction in-plane. Fig. 6 a-d shows the case of in-plane spin transfer torque which causes domain expansion into the spacer region without touching the corners[17,23]. When the current is turned off (Fig. 6 d) the stable domain configuration is that of a replication of the reversal region in the second constriction, similar to the situation shown in Fig. 5c-d. However, if the effective anisotropy points dominantly out of plane we find that the domain propagation into the spacer is gradual and even instantaneous decoupling of domains may occur. When the current is turned off, the stable configuration is one where the contraction of the domain back to the initial state is hindered by the sharp corners acting as pinning sites in the spacer region in Fig. 6e-h similar to the case of Fig. 5e-h. Therefore, the use of spin polarized current pulses results in either domain expansion to or replication in the next constriction along the direction of electron flow depending on the effective spin polarization direction. This implies that in a magnetic page memory device using only one read/write element per nanowire in conjunction with spin polarized current pulses to propagate the information is sufficient to generate the first storage layer which can be vertically replicated to the next layer on top. This domain injection is qualitatively similar to blowing of magnetic skyrmion bubbles[24].

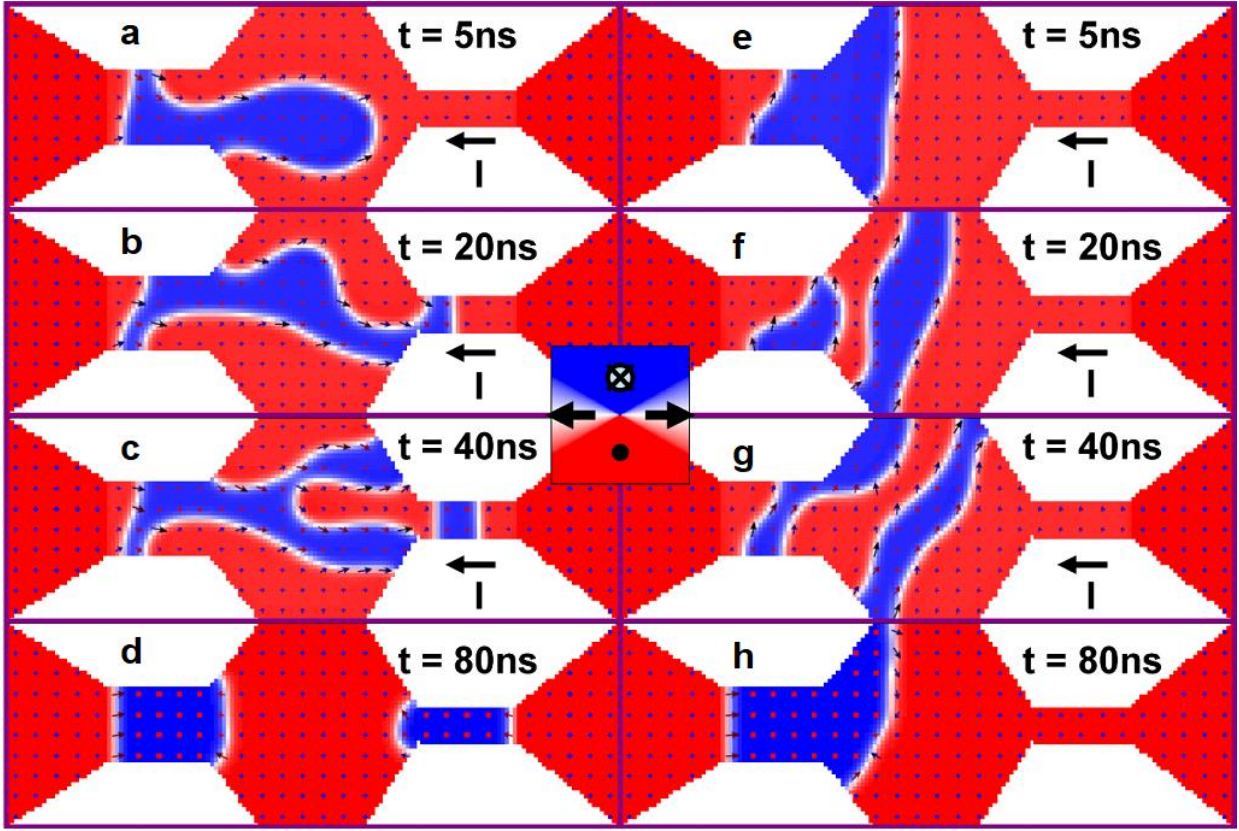

**Figure 6: Micromagnetic Simulations of Lateral Domain Transfer.**

**a, b and c,** The micromagnetic configuration of a double constriction sample (200nm and 100nm) at 5ns, 20 ns and 40 ns after a 4mA electrical current is turned on. **d,** The micromagnetic configuration 30ns after the current is turned off.
**e, f and g,** The micromagnetic configuration of a double constriction sample (200nm and 100nm) at 5ns, 20 ns and 40 ns after a 4mA electrical current is turned on assuming an out of plane polarization. **h** The micromagnetic configuration 30ns after the current is turned off. The color scale represents the z-component of the magnetization as shown at the center and the arrows represent the local magnetization direction.

In conclusion we have demonstrated the operation of a three dimensional magnetic page memory device that takes advantage of a combination of vertical data flow using local heating from current pulses in the presence of perpendicular stray fields and intra-layer horizontal flow using spin-polarized current pulses. Such an approach could lead to a memory solution fully utilizing three dimensional space with ultra-dense data storage capacity. The current work on the three dimensional magnetic page memory can also be potentially beneficial for the future applications of newly discovered complex magnetic nanostructures such as magnetic skyrmions, domain walls and solitons which could replace current magnetic bits with potentially higher mobilities due to interactions with spin polarized currents as well as the spin-orbit torque from pure spin currents.

**ACKNOWLEDGEMENTS**

This work was supported in part by the Bogazici University Research Fund under Grant number 16B03P5, the Scientific and Technological Research Council of Turkey TUBITAK under Contract number 112T205, the french PIA project "Lorraine Université d'Excellence", reference ANR-15-IDEX-04-LUE, CNRS/TUBITAK "PICS "program (2016). The authors acknowledge the help of Dr. Andreas Moser for magnetic force microscopy imaging and analysis, Dr. Jordan A. Katine for device nanofabrication, Dr. Jan-Ulrich Thiele for optimization of materials, Dr. Sylvia H. Florez for magnetic characterization and Dr. Bruce D. Terris for interpretation of the data.

[18] Jaafar, M., Asenjo, A., and Vazquez, M. Calibration of Coercive and Stray Fields of commercial magnetic force microscope probes *IEEE Trans. On Nanotech.* **7**, 245 (2008).

[19] Ozatay, O., Chalsani, P., Emley, N. C., Krivorotov, I. N., Buhrman, R. A. Magnetoresistance and magnetostriction effects in ballistic ferromagnetic nanoconstrictions *J. Appl. Phys.* **95**, 7315-7317 (2004).

[20] Marrows, C. Spin-polarized currents and magnetic domain-walls *Advances in Physics* **54,** 585-713 (2005).

[21] Kilic, U. et al. Magnetic switching driven by nanosecond scale heat and magnetic field pulses: An application of macrospin Landau-Lifshitz-Bloch model *Appl. Phys. Lett.* **101**, 252407 (2012).

[22] Atxitia, U., Hinzke, D. and Nowak, U. Fundamentals and applications of the Landau-Lifshitz-Bloch equation *J. Phys. D: Appl. Phys.* **50**, 033003 (2017).

[23] Claudio-Gonzalez D., Thiaville A., and Miltat J. Domain Wall Dynamics under Nonlocal Spin-Transfer Torque *Phys. Rev. Lett.* **108,** 227208 (2012).

[24] Jiang, W. et al. Blowing magnetic skyrmion bubbles *Science* **349,** 283-286 (2015).

# SUPPLEMENTARY NOTES

**Supplementary Note 1: Finite Element Simulations of the Temperature Distribution**

To ensure the reliable operation of a magnetic page memory array it is important to gain a full understanding of the nature of heat flow to the neighboring elements in the vertical stack. While it is beneficial to have good thermal isolation between memory elements to reduce the current levels required to push the coercivity down to the appropriate levels for writing, it is also important to have an efficient thermal dissipation mechanism so that the thermal relaxation occurs much faster than the operation speed of the device. The failure to have a fast enough thermal relaxation process would imply excessive temperature build-up between switching events ultimately leading to device failure due to the appearance of thermal instability of the magnetization and/or modification of the magnetic properties due to thermally driven interdiffusion as the instantaneous element temperature increases well beyond the Curie temperature. For this reason a detailed characterization of the thermal conductivities of individual layers and interface in the page memory stack is necessary.

The bulk thermal conductivity of the [CoNi]/Pd stack was measured using picosecond thermoreflectance technique at room temperature to be[1] 130 MW/m.K with no measurable temperature dependence up to 220ºC. The spacer between the elements is a 20 nm thick silicon-nitride layer with a measured ~3.2W/m.K bulk thermal conductivity at room temperature. However the interfacial thermal conductivity between silicon-nitride and [CoNi]/Pd stack which dominates the heat flow in a nanoscale page memory device was measured[2] to be 0.026W/m.K, yielding more than two orders of magnitude reduction in the thermal conductivity. Such high resistance to heat flow at the interface between a metal and an insulator is an indication of a poor overlap between the phonon states providing few phonon modes available for heat transfer[3]. At 220ºC it was found that the interfacial thermal conductivity displays a modest increase by 10% to ~0.03 W/m.K.

We have implemented a fully three dimensional finite element simulations of the heat flow between a 100 nm wide page memory element and its neighbor . The resulting thermal profile at the end of a write pulse is as shown in Supplementary Fig. 1. When the element of interest reaches the temperature necessary for the write operation (~110 ºC) as in Supplementary Fig. 1a, the neighboring element (the bottom element) reaches a maximum temperature of ~60 ºC far below the writing temperature as in Supplementary Fig. 1b. Therefore while it is clear that there is some thermal disturbance during the write process, this would cause a reduction of coercivity only from 3kOe to 2kOe[4] still implying very good magnetic stability.

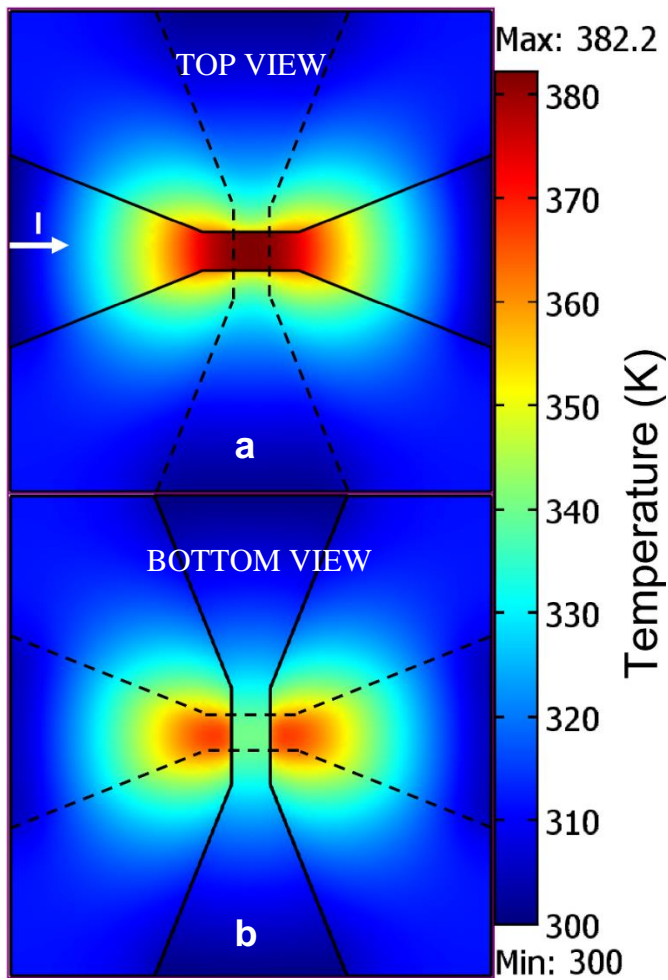

**Supplementary Figure 1: Finite element simulations of the temperature distribution profile in a magnetic page memory junction**. **a,** The top-view of the temperature distribution at the end of a 10ns current pulse applied to the top nanowire. The solid lines represent the boundaries of the top nanowire and the dotted lines mark the position of the bottom nanowire. **b,** The bottom-view of the temperature distribution at the end of a 10ns current pulse applied to the top nanowires- the thermal disturb to the bottom nanowire. The solid lines represent the boundaries of the bottom nanowire and the dotted lines mark the position of the top nanowire.

## Supplementary Note 2: Magnetic Force Microscopy Studies of Stray Field Induced Writing

In order to demonstrate the successful write operation on a single magnetic wire conclusively, the electrical setup for nanosecond pulse injection was prepared on the metallic (grounded) sample stage of a Veeco Dimension 3100 Atomic Force Microscope (AFM) system. The magnetic wires (made of perpendicular anisotropy CoNi/Pd multilayers) were wire bonded onto a grounded chip carrier. The wire bonds were placed carefully to avoid any potential contact with the MFM tip in the subsequent writing and imaging experiments. Before microscope measurements, all wires were magnetized to saturation by applying a 2T perpendicular to plane magnetic field in an electromagnet. Therefore, the initial magnetic state of each wire was the remnant state with close to 100% out of plane magnetic orientation.

The magnetic wires were then placed under the microscope and electrically connected to a Picosecond Pulse Labs 10000 series programmable pulse generator and a Keithley 2400 Sourcemeter via a bias tee. A magnetic force microscope tip (Bruker MESP-HM) was brought to 1nm distance above the surface of the magnetic wire by focusing on a 50 by 50 nm$^2$ area and slowing down the scan rate to 0.01Hz during the lift-mode scan which practically parked the tip right above the surface. The wire resistance was monitored with the sourcemeter before and after the application of electrical pulses. The wire resistances varied between 50 Ω to kΩ for wires between 2μm wide down to 100nm wide. Any reflected pulse due to impedance mismatch was absorbed by the 50 Ω output impedance of the pulser. Supplementary Fig. 2 shows the magnetic images of a 2 μm wide wire before writing and after the application of a series of 7.4 ns 3.2 V pulses in the presence of the stray field from a magnetic tip. A dot pattern Supplementary Fig. 2b and a word (HITACHI) Supplementary Fig. 2c were successfully written with the aforementioned method using the same pulse parameters.

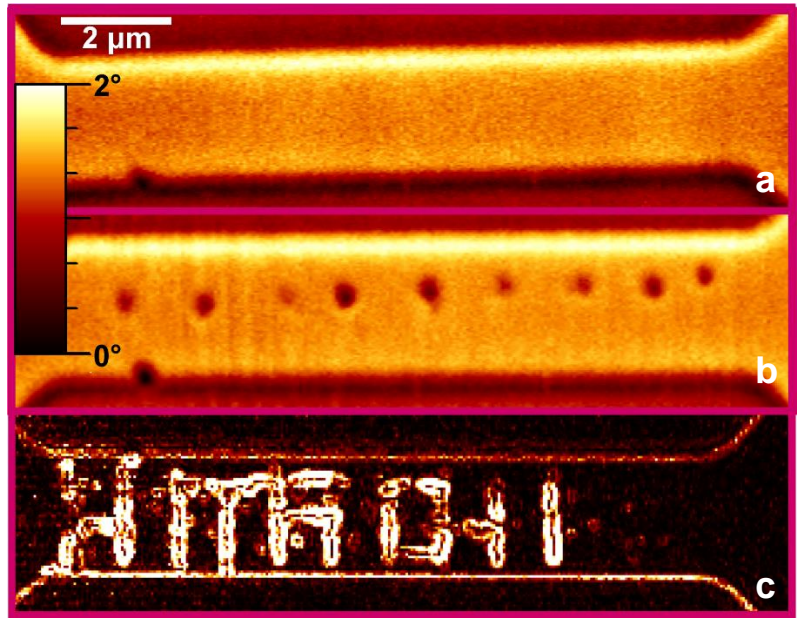

**Supplementary Figure 2: Magnetic force microscopy (MFM) studies of writing onto [CoNi]/Pd microwires with a magnetic tip. a,** The MFM image of the remanent state after saturation with a 2T out of plane applied field. **b,** The MFM image after writing a dot pattern with 7.4 ns 3.2V pulses ( ~ $3.10^7$ A/cm$^2$ )- 1nm tip height **c,** The MFM image after writing HITACHI by controlling tip position with the same pulse parameters as in (b). The color scale is inverted for more clarity.

## Supplementary Note 3: Magnetic Characterization

A full magnetic characterization of the bulk [$Co_{55}Ni_{45}$ (0.22) / Pd (1.2)]$_{22}$ perpendicular anisotropy magnetic thin films was carried out as a function of temperature with an ADE Microsense Model 10 VSM with the heater option. Temperature dependence of perpendicular magnetic field hysteresis loops, coercive field and saturation magnetization of the full film were measured from room temperature up to the Curie temperature of 175 °C as shown in Supplementary Fig. 3a, b and d respectively together with an in-plane hysteresis loop measurement at 125 °C in Supplemetary Fig. 3c. It can be seen that at 125 °C the nucleation field for the full film drops down to zero together with the coercive field implying that at this temperature, in the absence of an applied magnetic field (i.e no Zeeman energy), the demagnetization energy exceeds the exchange energy causing a magnetization relaxation by breaking into domains of perpendicular orientation.

Supplementary Fig. 3d indeed shows that even though the Curie temperature appears to be at 175 °C as measured with 1kOe magnetic field applied perpendicular to the film plane, the magnetization goes through a gradual relaxation above 100 °C reaching the zero point at 125 °C. The perpendicular anisotropy field ($H_k$) measured from in-plane hysteresis loops reveals that the room-temperature $H_k$ measurement of 18 kOe drops to about 8 kOe (as shown in Supplementary Fig. 3c) at 125 °C. Since a substantial average out-of-plane anisotropy is still present it becomes clear that at 125°C the magnetization relaxation results in demagnetization by breaking into domains with perpendicular orientation.

The temperature range where the magnetic multilayer becomes susceptible to reorientation from the stray field of a magnetic tip can therefore be determined to be in the 110-125 °C range where the coercive field drops to near zero.

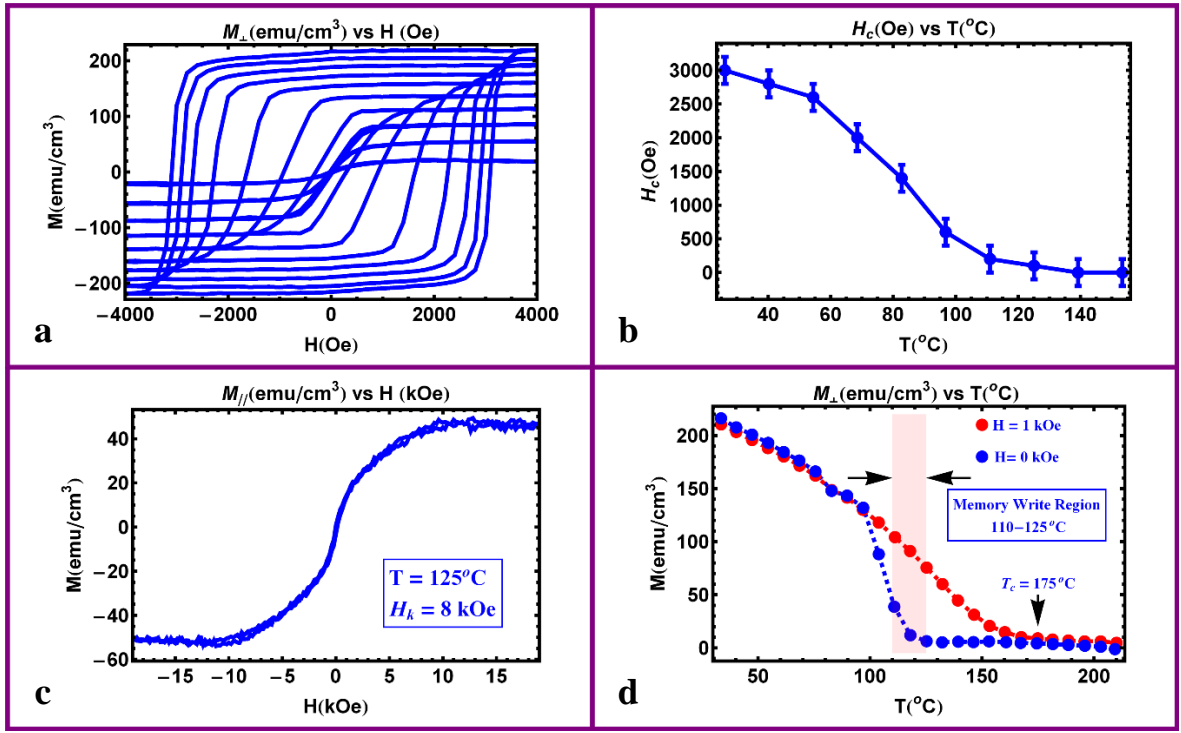

**Supplementary Figure 3: VSM measurements of the magnetic properties of a perpendicular anisotropy [Co$_{55}$Ni$_{45}$ (0.22) / Pd (1.2)]$_{22}$ magnetic multilayer thin film.**

   **a,** Magnetization (M) vs. magnetic field (H) applied perpendicular to the film plane hysteresis loops measured at temperatures from 25 °C to 150 °C.

   **b,** The coercive fields (zero crossing points) extracted from M vs H measurements shown in a as a function of temperature from 25 °C to 150 °C.

   **c,** In-plane hard axis loop measured at 125 °C.

   **d,** Magnetization vs. temperature measured in the presence of a 1kOe out-of-plane magnetic field (red dots) and in the absence of any applied magnetic field (blue dots).

## SUPPLEMENTARY METHODS

### Supplementary Method1: Sample Fabrication

We began the device fabrication by sputter depositing the bottom ferromagnetic layer, Ta(25A)/Pd(50A)/[CoNi(2.2A)/Pd(12A)] x 22/Pd(25A)/Ta(50A). The Ta layer served as a capping layer protecting the film during processing. Using a combination of e-beam lithography and ion milling, we patterned the bottom FM into the nanowires, with leads extending out far enough to allow electrical contact in a subsequent processing step. Following the ion mill, the devices were encapusulated in aluminum oxide, and chemical mechanical polishing (CMP) was used to create planar structures in which the top of the FM layer was coplanar with the insulating oxide.

Following planarization, $SiN_x$ of 20, 40, or 60 nm was sputtered onto the wafers. An *in situ* sputter clean removed much of the protective Ta cap prior to deposition. After the nitride deposition, the second ferromagnetic layer was sputter deposited. Using the same techniques used for the bottom FM layer, the top FM layer was patterned into nanowires orthogonal to the first layer's nanowires. As aluminum oxide and silicon nitride now covered the electrical leads to the bottom nanowires, contact vias were etched through these insulators, and the devices were completed by patterning Cr/Au contacts to the top and bottom FM nanowires.

## Supplementary Method 2: Landau-Lifshitz Bloch Macrospin Simulations of Thermally Assisted Stray Field Induced Writing

Since the bit transfer process in the transverse direction involves heating of the magnetic layers close to their Curie temperatures in order to momentarily decrease the coercivity and become susceptible to stray fields, the conventional Landau-Lifshitz-Gilbert (LLG) formalism is not applicable[5]. The rapid temperature change induced longitudinal fluctuations of the magnetization magnitude leading to a temperature dependent longitudinal damping term as a result of the dynamics of magnetic moments can be described with the Landau-Lifshitz Bloch (LLB) equation[5].

The LLB formalism properly accounts for the time evolution of the magnetization $M$ at elevated temperatures such that:

$$\frac{\partial \boldsymbol{m}}{\partial t} = -\tilde{\gamma}(\boldsymbol{m} \times \boldsymbol{H}_{eff}) + \frac{\tilde{\gamma}\alpha_{\parallel}}{|\boldsymbol{m}|^2}[\boldsymbol{m}\cdot(\boldsymbol{H}_{eff})]\boldsymbol{m} - \frac{\tilde{\gamma}\alpha_{\perp}}{|\boldsymbol{m}|^2}\boldsymbol{m}\times[\boldsymbol{m}\times(\boldsymbol{H}_{eff})]$$

where $\boldsymbol{m} = \frac{M}{M_s(T=0K)}$; $\tilde{\gamma}$ is the absolute value of the gyromagnetic ratio; $\boldsymbol{H}_{eff}$ is the effective magnetic field; $\alpha_{\parallel}$ and $\alpha_{\perp}$ are the longitudinal and transverse damping parameters respectively; $\boldsymbol{\zeta}_{\parallel}$ and $\boldsymbol{\zeta}_{\perp}$ are the longitudinal torque and transverse effective field terms due to Gaussian stochastic thermal processes. The temperature dependence of the longitudinal and transverse damping parameters is given by:

$$\alpha_{\parallel} = \alpha_G\left(\frac{2T}{3T_c}\right); \quad \alpha_{\perp} = \begin{cases} \alpha_G\left(1 - \frac{T}{3T_c}\right) & T < T_c \\ \alpha_G\left(\frac{2T}{3T_c}\right) & T \geq T_c \end{cases}$$

where $\alpha_G$ is the Gilbert damping parameter and $T_c$ is the Curie temperature. The effective field consists of the following components:

$$H_{eff} = H_{appl} + H_{an} + \begin{cases} \frac{1}{2\chi_\parallel}\left(1 - \frac{m^2}{m_e^2}\right)m & T \leq T_c \\ \frac{1}{2\chi_\parallel}[\frac{3}{5}\left(\frac{T_c}{T_c - T}\right)m^2 - 1]\,m & T \geq T_c \end{cases}$$

where $H_{appl}$ is the applied field, $H_{an}$ is the anisotropy field and the last term is the longitudinal effective field which depends on temperature T, longitudinal susceptibility of the material $\chi_\parallel$ and zero field equilibrium magnetization[6] $m_e$.

The simulation parameters are $\alpha_G = 0.1$, $M_S(T=300K)= 218\times10^3$A/m, the saturation magnetization dependence on temperature is obtained from Supplementary Fig. 3d. The anisotropy field is computed as $H_{an} = -D_X m_X - D_Y m_Y - D_Z m_Z$ where the out of plane demagnetizing tensor $D_Z$ is -0.1 and it takes into account either the shape demagnetizing field and the perpendicular anisotropy.

## Supplementary Method 3: Landau-Lifshitz Gilbert (LLG) Micromagnetic Simulations of Lateral Bit Transfer

Micromagnetic simulations were performed with the publicly available OOMMF code[7] to study the spin transfer torque induced lateral transport of bits between consecutive constrictions. Temperature dependent material parameters obtained from magnetometry measurements as shown in Supplementary Fig. 3 were used. For the applied current pulses, the heating effect was taken into account by considering the temperature dependence of magnetization and anisotropy.

## **Supplementary References**